\documentclass[a4paper,11pt]{article}
\usepackage{wrapfig}

\usepackage{pos}
\usepackage{lipsum} 
\usepackage{lineno}

\usepackage{booktabs}
\usepackage{siunitx}
\title{Analysis of the Diffuse Astrophysical Flux from the Galactic Plane with 12.1 Years of IceCube Starting Tracks, Throughgoing Tracks and Cascades}

\ShortTitle{Analysis of the Galactic and Extragalactic neutrino flux with DNN Cascades, ESTES and Northern Tracks}

\author{The IceCube Collaboration \\{\normalsize \normalfont(a complete list of authors can be found at the end of the proceedings)}\\}

\emailAdd{jsosborn2@wisc.edu, karle@icecube.wisc.edu}

\abstract{

The IceCube Neutrino Observatory has measured an isotropic astrophysical neutrino flux through various detection channels for over 12 years. IceCube has also detected neutrino emission from the Galactic plane at the 4.5$\sigma$ significance level compared to a background-only hypothesis, testing three models of Galactic diffuse emission: Fermi-LAT $\pi^0$, KRA$_{\gamma}^{5}$, KRA$_{\gamma}^{50}$. We present an analysis combining 3 detection channels: throughgoing tracks, starting tracks and cascades. The throughgoing track sample is restricted to the northern sky to reduce atmospheric backgrounds, while the starting track and cascade samples reduce the atmospheric neutrino backgrounds in the southern sky by vetoing accompanying muons. We will use this combination of event samples from 12.1 years of data to measure the Galactic neutrino spectrum in the TeV to PeV energy range and independently for multiple galactic regions in a model independent procedure. We will simultaneously measure the isotropic cosmic neutrino flux.

\vspace{4mm}

{\bfseries Corresponding authors:}
Jesse Osborn$^{1*}$, Albrecht Karle$^{1}$\\
{$^{1}$ \itshape University of Wisconsin, Madison}\\[4mm]
$^*$ Presenter
}

\ConferenceLogo{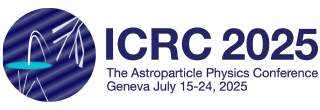}

\FullConference{39th International Cosmic Ray Conference (ICRC2025)\\
 15–24 July 2025\\
Geneva, Switzerland\\}

\begin{document}

\maketitle

\section{Introduction}\label{1}

IceCube is a cubic-kilometer neutrino detector located at the South Pole. IceCube has detected neutrino emission from the Galactic plane with high significance \cite{DNN_Cascades_Science_2023}, establishing it as a subdominant component of the diffuse astrophysical flux which has been characterized by IceCube for years. Studying this Galactic neutrino flux can help with characterizing the isotropic astrophysical neutrino flux in the TeV energy range where the Galactic component may be a significant contribution to the total astrophysical flux. It can also help to better understand cosmic ray interactions and acceleration within our galaxy which is presumed to be responsible for the Galactic neutrino emission.

We present here the initial stages of an analysis to better characterize the spatial and spectral features of Galactic neutrino emission using 12.1 years of IceCube cascade, starting track and track data. IceCube has ongoing analyses \cite{Jonas_2025icrc} that make use of theoretical neutrino emission models such as Fermi-LAT $\pi^0$ \cite{Fermi_pi0}, KRA$_{\gamma}^{5}$ / KRA$_{\gamma}^{50}$ \cite{KRA_gamma} or CRINGE \cite{CRINGE} in the context of diffuse flux measurements. In contrast, this analysis does not rely on any theoretical emission models. This analysis works towards making an independent measurement of Galactic neutrino emission using generic spatial assumptions which will allow for a more robust study of the spatial and spectral features.

\section{Event Selections}\label{2}

\begin{figure}[h!]
    \centering
    \includegraphics[width=0.32\textwidth]{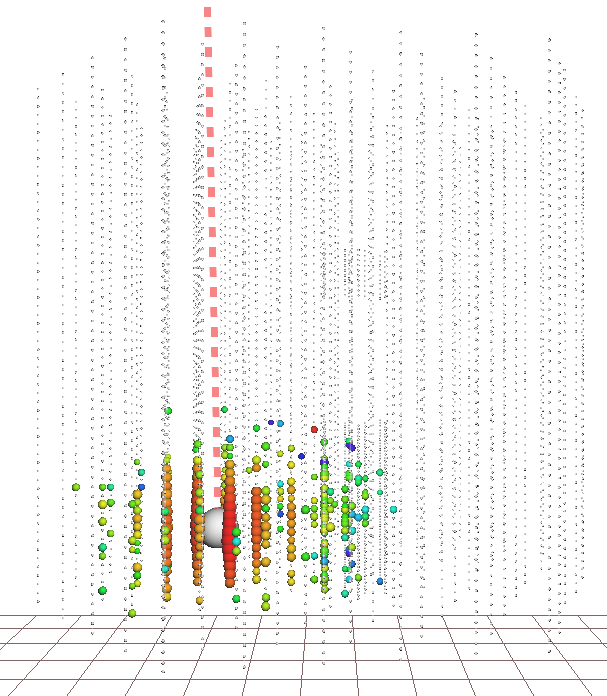}
    \includegraphics[width=0.32\textwidth]{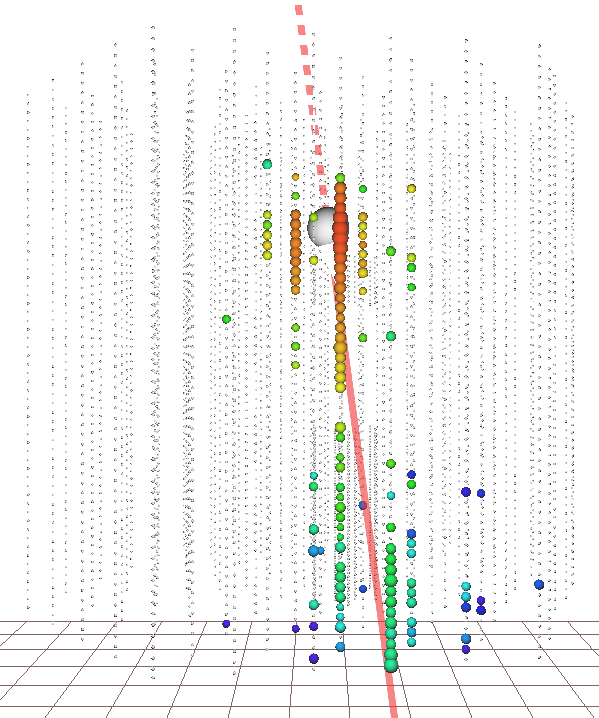}
    \includegraphics[width=0.32\textwidth]{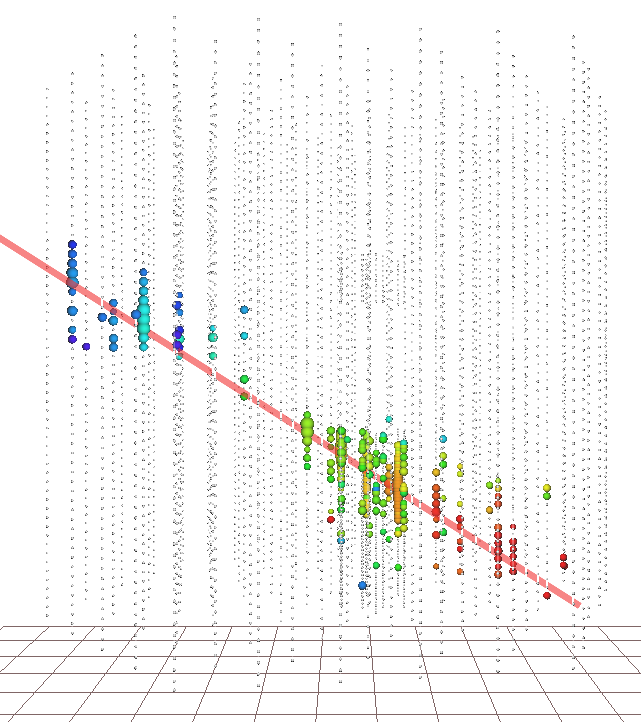}
    \caption{Displays of final level Monte Carlo events for each of the three selections. Red represents the earliest light detected for the event, while blue represents the last light detected. Where applicable, a dashed red line denotes the trajectory of a neutrino while a solid red line denotes the trajectory of a muon, and a large gray sphere denotes the interaction vertex of a neutrino. Left: A DNN Cascades 18 TeV charged current $\nu_e$ cascade event from the southern sky. Middle: An ESTES 22 TeV charged current $\Bar{\nu_{\mu}}$ starting track event from the southern sky. Right: A Northern Tracks 77 TeV charged current $\nu_{\mu}$ throughgoing event from the northern sky.}
    \label{fig:example event views}
\end{figure}

To achieve the best possible sensitivity to a neutrino flux from the Galactic plane, this analysis makes use of a combination of three event selections each targeting a different event morphology: cascades, starting tracks and throughgoing tracks. Examples of events from each of the three selections can be seen in Fig. \ref{fig:example event views}.

\subsection{Deep Neural Network (DNN) Cascades}\label{sec2.1}

DNN Cascades makes use of deep learning and neural networks to retain cascade events at lower energy thresholds and nearer to or slightly outside the boundary of the instrumented volume of the detector compared to previous cascade selections. Since the selection is tuned for the cascade event morphology (all neutral-current $\nu$ interactions as well as charged-current $\nu_{e}$ and $\nu_{\tau}$ interactions), it has the best all flavor sensitivity, great energy resolution and high statistics. The selection uses the self-veto effect \cite{self-veto} to obtain reasonable astrophysical purity in the southern sky where the majority of the Galactic plane resides. A variation of this event selection was previously used to detect Galactic neutrino emission at the level of 4.5$\sigma$ \cite{DNN_Cascades_Science_2023} which motivated this analysis. The variation of the selection used here has been updated and optimized for diffuse astrophysical flux measurements \cite{Zoe_2025icrc}.

\subsection{Enhanced Starting Track Event Selection (ESTES)}\label{sec2.2}

ESTES makes use of a dynamic starting track veto as well as machine learning algorithms to search for events with vertices contained within the detector volume and exiting muon tracks. These types of events are induced by charged current $\nu_{\mu}$ interactions within the detector. The exiting track gives the selection a good angular resolution while the contained vertex gives the selection a good energy resolution. ESTES also leverages the self veto effect \cite{self-veto} for a strong rejection of atmospheric events in the southern sky where the majority of the Galactic plane resides. This selection was previously used for an isotropic diffuse astrophysical flux measurement \cite{ESTES_PRD_2024} with sensitivity down to 3 TeV.

\subsection{Northern Tracks}\label{sec2.3}

By restricting itself to events from the northern sky (through the Earth's bedrock with respect to IceCube) and making use of machine learning algorithms, Northern Tracks has a high neutrino purity while also maintaining high statistics. Since the selection is tuned for throughgoing tracks, it has the best angular resolution of the three samples used here. However, being restricted to the northern sky limits its sensitivity to the Galactic plane (whose center is in the southern sky), but the high statistics nature of the sample still gives it considerable leverage. This selection was also previously used for an isotropic diffuse astrophysical flux measurement by itself \cite{Northern_Tracks_APJ_2022} and as a part of a combined fit alongside a different cascade event selection \cite{Combined_Fit_ICRC_2023}. The version used here will include an update to the energy reconstruction \cite{Emre_2025icrc}.

\section{Analysis Method}\label{3}

Each event selection contains atmospheric muons, atmospheric neutrinos (conventional and prompt), astrophysical neutrinos from the isotropic diffuse flux and astrophysical neutrinos from the Galactic plane. The Monte Carlo (MC) is given an expected number of events $\lambda$ for each of these:

\begin{equation*}
\lambda_{\mathrm{Total}} = \lambda_{\mathrm{Galactic\ Astro}} + \lambda_{\mathrm{Iso\ Astro}} + \lambda_{\mathrm{Conv}} + \lambda_{\mathrm{Prompt}} + \lambda_{\mathrm{Muon}}.
\label{eq:lambda}
\end{equation*}

The atmospheric components ($\lambda_{\mathrm{Conv}}$, $\lambda_{\mathrm{Prompt}}$ and $\lambda_{\mathrm{Muon}}$) are modeled by scaling normalization factors of flux model components, namely the Gaisser H4a cosmic-ray model \cite{h4a} and Sibyll 2.3c hadronic interaction model \cite{23c}. The isotropic astrophysical component is modeled here assuming a single power law flux, although this analysis will also test other isotropic models such as a broken power law or log parabola as has been done in previous analyses \cite{ESTES_PRD_2024}. In the work shown in this proceeding, the three selections are still treated independently and are given different parameter assumptions based on previous measurements.
\begin{enumerate}
    \item For DNN Cascades, the assumed isotropic astrophysical flux is the single power law fit of the previous combined fit measurement using Northern Tracks and a different cascade sample \cite{Combined_Fit_ICRC_2023} ($\phi_{\mathrm{Iso\ Astro}} = 1.80$, $\gamma_{\mathrm{Iso\ Astro}} = 2.52$) although a diffuse measurement using DNN Cascades for the first time is currently underway \cite{Zoe_2025icrc}. All other systematics are nominal.
    \item For ESTES, the assumed isotropic astrophysical flux as well as the other systematic parameters come from the 10.3 year measurement \cite{ESTES_PRD_2024} ($\phi_{\mathrm{Iso\ Astro}} = 1.68$, $\gamma_{\mathrm{Iso\ Astro}} = 2.58$).
    \item For Northern Tracks, the assumed isotropic astrophysical flux is that of the Northern Tracks paper \cite{Northern_Tracks_APJ_2022} ($\phi_{\mathrm{Iso\ Astro}} = 1.44$, $\gamma_{\mathrm{Iso\ Astro}} = 2.37$). All other systematics are nominal.
\end{enumerate}

\begin{figure}[h!]
    \centering
    \includegraphics[width=0.48\textwidth]{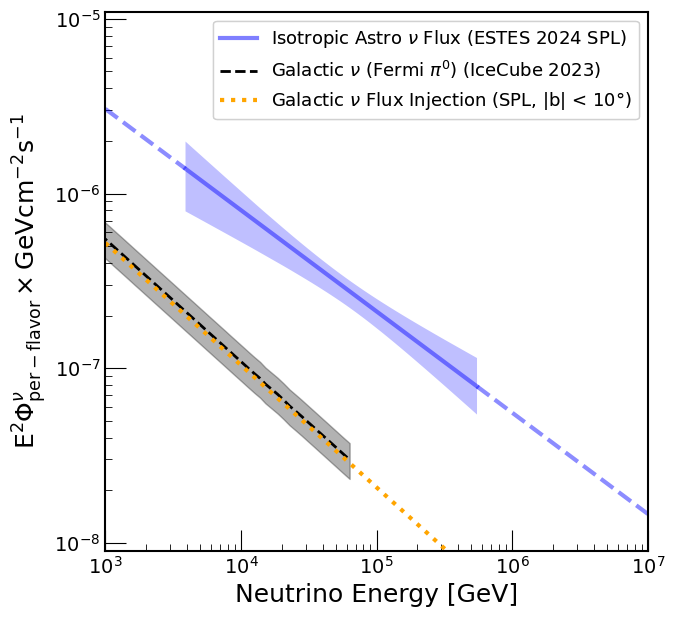}
    \caption{Energy scaled, all sky, per-flavor neutrino flux plot. The blue line and shaded band is the 10.3 year ESTES measurement of the isotropic astrophysical flux \cite{ESTES_PRD_2024}. The dashed gray line and shaded band are the best fit Galactic neutrino flux using the Fermi-$\pi^0$ template from the 2023 IceCube Science result using DNN Cascades \cite{DNN_Cascades_Science_2023}. The dotted orange line is injected to the Galactic plane band shown in the left of Fig. \ref{fig:skymap diagrams} to test this analysis's sensitivity.}
    \label{fig:flux plot}
\end{figure}

The Galactic plane astrophysical component is also modeled assuming a single power law flux:
\begin{equation*}
\begin{split}
&\Phi_{Astro\_GP}^{Total} = \phi_{\mathrm{Astro\_GP}}^{\mathrm{per-flavor}} \times (\frac{\mathrm{E}_{\nu}}{100 \mathrm{TeV}})^{-\gamma_{GP}} \times \mathrm{C}_{0} \hskip0.5em (|b| < 10^{\circ}),\\
&\ \hskip6em 0 \hskip12.6em (|b| \geq 10^{\circ}),\\
&\mathrm{where}, \hskip0.1em \mathrm{C}_{0} = 3 \times 10^{-18} \times \mathrm{GeV}^{-1} \hskip0.1em \mathrm{cm}^{-2} \hskip0.1em \mathrm{s}^{-1} \hskip0.1em \mathrm{sr}^{-1}, \hskip0.3em \phi_{\mathrm{Astro\_GP}}^{\mathrm{per-flavor}} = 0.95, \hskip0.3em \gamma_{GP} = 2.7.\\
\label{eq:SPL}
\end{split}
\end{equation*}
This injection flux ($\phi_{\mathrm{Galactic\ Astro}} = 0.95$, $\gamma_{\mathrm{Galactic\ Astro}} = 2.7$) is derived by setting the all sky integrated emission from our Galactic box equal to the all sky integrated emission that was measured by the IceCube in 2023 using the Fermi-LAT $\pi^0$ template with DNN Cascades \cite{DNN_Cascades_Science_2023}. A comparison of the fluxes can be seen in Fig. \ref{fig:flux plot}.

\begin{figure}[h!]
    \centering
    \includegraphics[width=0.49\textwidth]{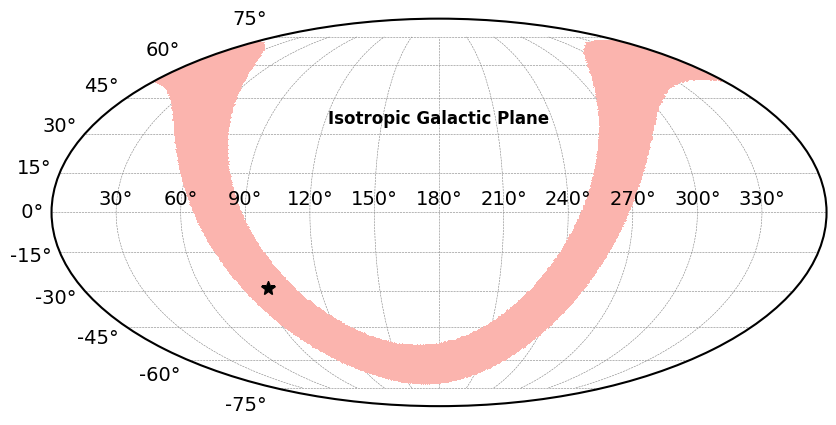}
    \includegraphics[width=0.49\textwidth]{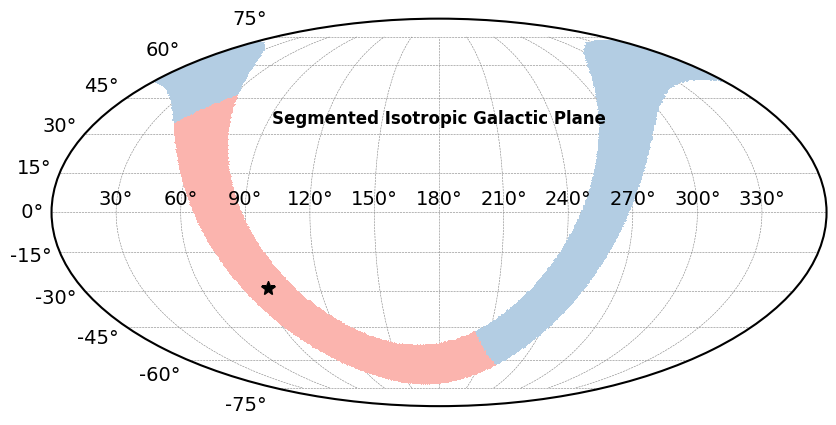}
    \caption{Left: A sky map illustrating a band (|b| < 10$^{\circ}$) surrounding the Galactic plane. Sensitivity to single power law emission from this band is shown in Fig. \ref{fig:sensitivity plots}. Right: A sky map illustrating two bands (|b| < 10$^{\circ}$, l = (-80$^{\circ}$, +80$^{\circ}$) and l = 
    (-180$^{\circ}$, -80$^{\circ}$), (+80$^{\circ}$, +180$^{\circ}$)). This analysis will also probe for spatial structure in Galactic neutrino emission by testing for emission from various segmented structures.}
    \label{fig:skymap diagrams}
\end{figure}

The spatial extension of the Galactic plane is modeled assuming a generic rectangular box in galactic coordinates (galactic latitude < 10$^{\circ}$) as shown in the left panel of Fig. \ref{fig:skymap diagrams}. The extension of the box in galactic latitude will be optimized in this analysis. This analysis also plans to test other spatial extensions, such as segmenting the Galactic plane into multiple boxes as seen in the right panel of Fig. \ref{fig:skymap diagrams}. This would allow us to test whether we see more emission from the inner or outer regions of the galaxy. This analysis will strive to balance the amount of spatial segments with the sensitivity to spectral features in each spatial segment to characterize the Galactic neutrino emission as finely as possible both spatially and spectrally.

\begin{figure}[h!]
    \centering
    \includegraphics[width=0.99\textwidth]{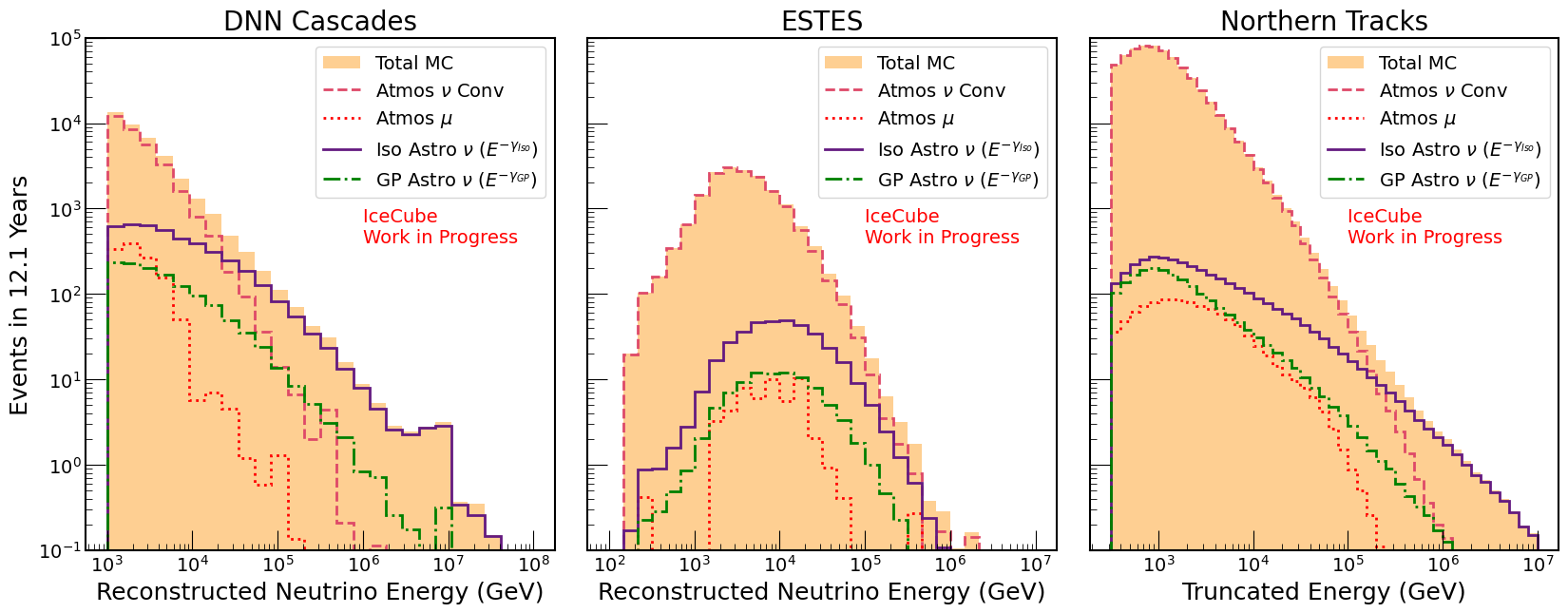}
    \caption{The reconstructed energy distributions for each of the three selections. Here the MC is weighted using pre-fit parameters based on earlier measurements and baseline values.}
    \label{fig:energy hists}
\end{figure}

Using pre-data-fit parameter assumptions, we plot the reconstructed energy distributions of our selections in Fig. \ref{fig:energy hists}. The Galactic neutrino emission, even more so than the isotropic astrophysical emission, is a subdominant signal compared to the atmospheric backgrounds in each selection. We expect different spectral features between the isotropic astrophysical emission, Galactic neutrino emission and atmospheric emission which is why we use reconstructed energy as an observable.

\begin{figure}[h!]
    \centering
    \includegraphics[width=0.99\textwidth]{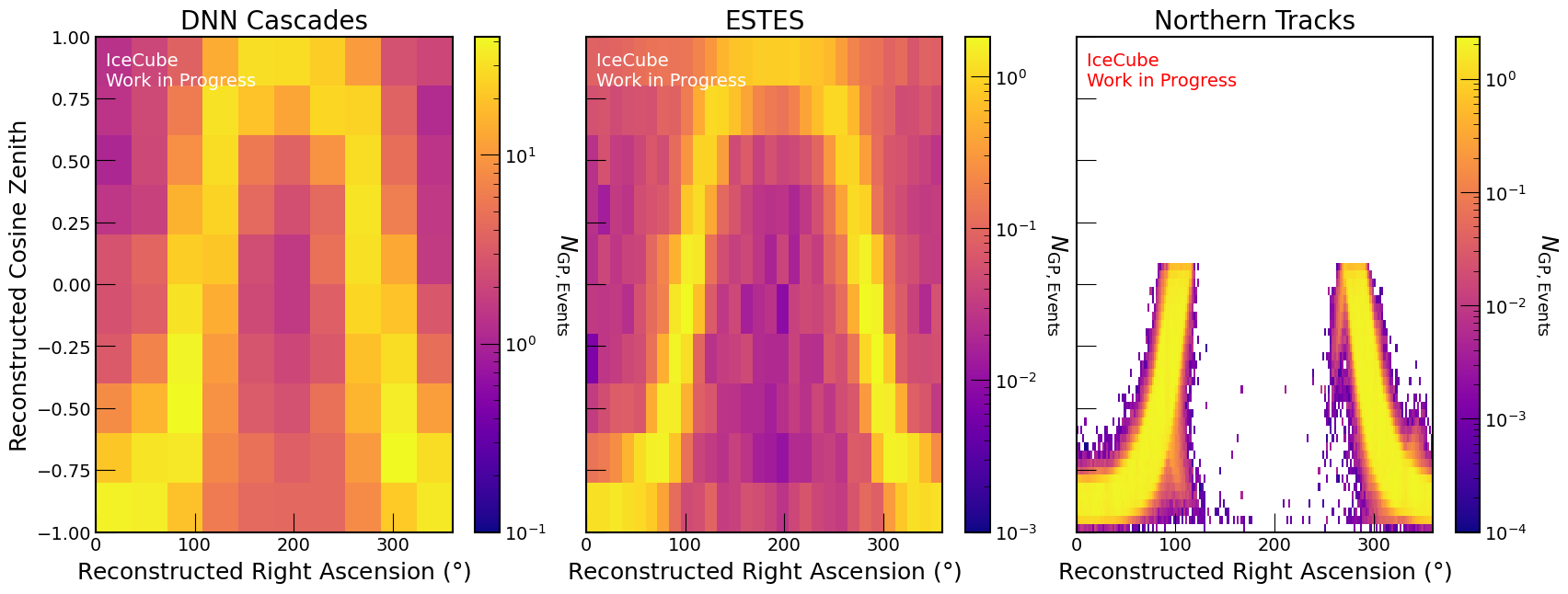}
    \caption{The reconstructed cosine zenith vs reconstructed right ascension distributions for each of the three selections. The color axes denotes the number of Galactic plane neutrinos in each bin. Here the MC is weighted using pre-fit parameters based on earlier measurements and baseline values.}
    \label{fig:cos vs ra hists}
\end{figure}

We also plot the cosine zenith vs right ascension distributions of our Galactic neutrino emission for each selection in Fig. \ref{fig:cos vs ra hists}. These distributions demonstrate the spatial features of Galactic neutrino emission which are essential to this analysis's sensitivity. This why the analysis needs to make use of the reconstructed cosine zenith and right ascension as observables. Better angular resolution allows for finer binning in these observables which allows for better capturing of the spatial features and improving sensitivity.

Table \ref{tab1} shows the expected event rates using MC from each flux component for each selection in 12.1 years of IceCube data taking.
\begin{table}[h!]
\centering
\begin{tabular}{l
                S[table-format=6.1]@{\,}r
                S[table-format=5.1]@{\,}r
                S[table-format=7.1]@{\,}r}
\toprule
& \multicolumn{2}{c}{DNN Cascades} & \multicolumn{2}{c}{ESTES} & \multicolumn{2}{c}{Northern Tracks} \\
\midrule
GP Astro $\nu$       & 1269.5 & (3.19\%)   & 90.9    & (0.51\%)   & 1971.6  & (0.31\%) \\
Iso Astro $\nu$      & 4400.3 & (11.07\%)  & 373.4   & (2.10\%)   & 3452.7  & (0.54\%) \\
Atmos Conv $\nu$     & 32861.3 & (82.67\%) & 17237.8 & (97.10\%)  & 636073.4 & (98.99\%) \\
Atmos Conv $\mu$     & 1219.5 & (3.07\%)   & 51.3    & (0.29\%)   & 1061.4  & (0.17\%) \\
Total MC             & 39750.5 & (100.00\%) & 17753.4 & (100.00\%) & 642559.1 & (100.00\%) \\
\bottomrule
\end{tabular}
\caption{Pre-fit expected event rates in 12.1 years from MC for each of the three selections.}
\label{tab1}
\end{table}

The analysis will use a forward-folding binned likelihood fit in three dimensions: $N$ energy bins, $M$ cosine zenith bins and $O$ right ascension bins. The likelihood of observing $k$ events when expecting $\lambda$ events is modeled using a Poisson probability, and the negative log-likelihood function is minimized. The test statistic is defined as the negative log-likelihood ratio of the likelihood function with a set of parameters $\Theta$ with respect to the likelihood function with a set of parameters that best describe the data $\hat\Theta$:
\begin{equation*}
\mathcal{L}(\lambda | k) = \Pi ^{N \cdot M \cdot O}_{i=1} (\frac{{e^{ - \lambda_i } \lambda ^k }}{{k!}}), \hskip0.3em \mathcal{TS}(\Theta) = -2 \mathrm{log} (\frac{\mathcal{L}(\lambda(\Theta) | k)}{\mathcal{L}(\lambda(\hat\Theta) | k)} ).
\label{eq:llh}
\end{equation*}

\section{Sensitivity}\label{4}

\begin{figure}[h!]
    \centering
    \includegraphics[width=0.45\textwidth]{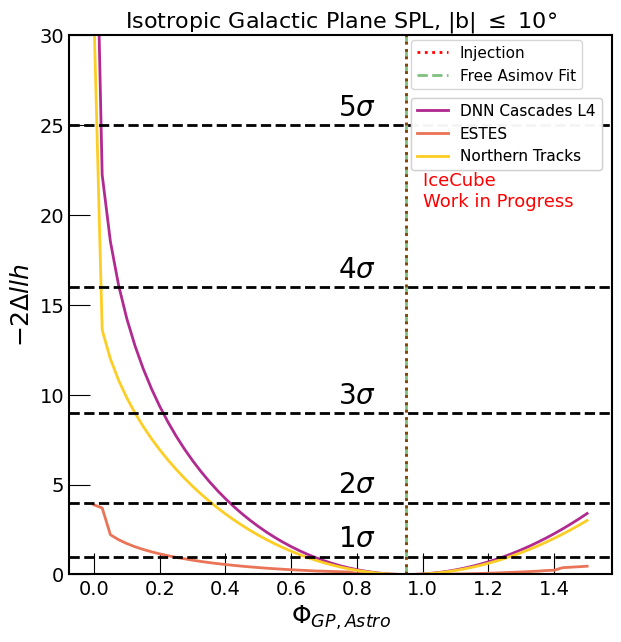}
    \includegraphics[width=0.45\textwidth]{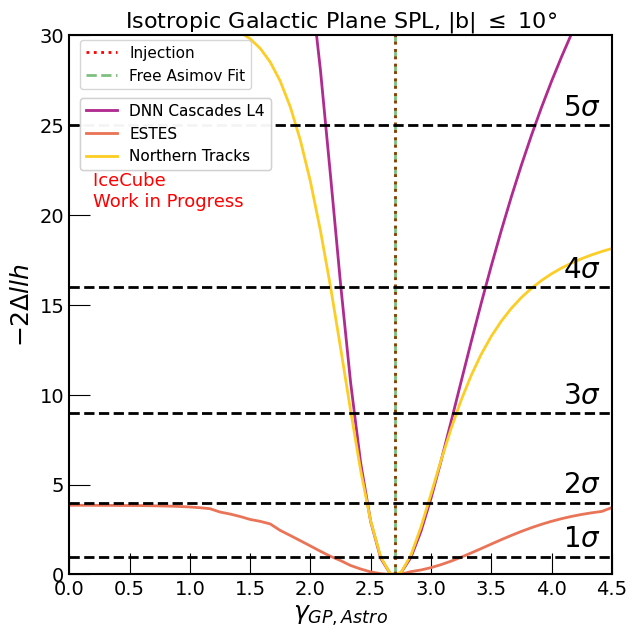}
    \caption{One dimensional log-likelihood scans for the normalization (left) and spectral index (right) of a single power law flux injected from the Galactic plane as seen in Fig. \ref{fig:flux plot}. The confidence lines are presented using Wilks' theorem.}
    \label{fig:sensitivity plots}
\end{figure}

In Fig. \ref{fig:sensitivity plots}, we show two 1 dimensional Asimov profile likelihood scans of the Galactic plane astrophysical flux parameters assuming a uniform single power law in a box of $\pm$ 10$^{\circ}$ in Galactic latitude, as shown in Figs. \ref{fig:flux plot} and \ref{fig:skymap diagrams}. The plot of the sensitivity to the normalization demonstrate the ability for this analysis to confirm Galactic neutrino emission at a high significance as IceCube has previously done \cite{DNN_Cascades_Science_2023}. The sensitivity to the spectral index demonstrates the novel ability of this analysis to characterize the spectral features of the Galactic neutrino emission, something which cannot be done using theoretical emission templates which have fixed spectral shapes.

The model independent nature of this analysis allows for more granular tests of Galactic neutrino emission. Spatially, the size of the box in galactic latitude can be increased or decreased to probe the extent of emission. The box can be split into multiple segments as in the right panel of Fig. \ref{fig:skymap diagrams} to test if there is greater emission coming from one segment of the galaxy relative to another. Spectrally, this analysis can probe using various analytical functions: a single power law, a broken power law, a log parabola or a segmented power law.

At the same time, this analysis will be measuring the spectral shape of the isotropic astrophysical neutrino flux using the same analytical functions with a novel combination of event selections. This allows for a robust understanding of the Galactic neutrino component and how it relates to the isotropic component of the astrophysical neutrino flux.

\section{Conclusion}\label{5}

This analysis intends to measure the Galactic neutrino spectrum using a procedure that is independent of the spatial and spectral constraints of theoretical models. This analysis will be performed using a combination of event selections targeting cascades, starting tracks and throughgoing tracks. In doing so, we hope to probe the spatial and spectral structure of the Galactic neutrino emission in order to inform our understanding of the emission independent of theoretical templates. This combination measurement will also serve a novel measurement of the isotropic component of the astrophysical neutrino flux.  

\bibliographystyle{ICRC}
\bibliography{references}

\clearpage

\section*{Full Author List: IceCube Collaboration}

\scriptsize
\noindent
R. Abbasi$^{16}$,
M. Ackermann$^{63}$,
J. Adams$^{17}$,
S. K. Agarwalla$^{39,\: {\rm a}}$,
J. A. Aguilar$^{10}$,
M. Ahlers$^{21}$,
J.M. Alameddine$^{22}$,
S. Ali$^{35}$,
N. M. Amin$^{43}$,
K. Andeen$^{41}$
% ...
\\
\\
% ...
$^{10}$ Universit{\'e} Libre de Bruxelles, Science Faculty CP230, B-1050 Brussels, Belgium \\
% ...
$^{16}$ Department of Physics, Loyola University Chicago, Chicago, IL 60660, USA \\
$^{17}$ Dept. of Physics and Astronomy, University of Canterbury, Private Bag 4800, Christchurch, New Zealand \\
% ...
$^{21}$ Niels Bohr Institute, University of Copenhagen, DK-2100 Copenhagen, Denmark \\
$^{22}$ Dept. of Physics, TU Dortmund University, D-44221 Dortmund, Germany \\
% ...
$^{35}$ Dept. of Physics and Astronomy, University of Kansas, Lawrence, KS 66045, USA \\
% ...
$^{39}$ Dept. of Physics and Wisconsin IceCube Particle Astrophysics Center, University of Wisconsin{\textemdash}Madison, Madison, WI 53706, USA \\
% ...
$^{41}$ Department of Physics, Marquette University, Milwaukee, WI 53201, USA \\
% ...
$^{43}$ Bartol Research Institute and Dept. of Physics and Astronomy, University of Delaware, Newark, DE 19716, USA \\
% ...
$^{63}$ Deutsches Elektronen-Synchrotron DESY, Platanenallee 6, D-15738 Zeuthen, Germany \\
$^{\rm a}$ also at Institute of Physics, Sachivalaya Marg, Sainik School Post, Bhubaneswar 751005, India \\
% ...

\subsection*{Acknowledgments}

\noindent
The authors gratefully acknowledge the support from the following agencies and institutions:
USA {\textendash} U.S. National Science Foundation-Office of Polar Programs,
U.S. National Science Foundation-Physics Division,
U.S. National Science Foundation-EPSCoR,
U.S. National Science Foundation-Office of Advanced Cyberinfrastructure,
Wisconsin Alumni Research Foundation,
Center for High Throughput Computing (CHTC) at the University of Wisconsin{\textendash}Madison,
Open Science Grid (OSG),
Partnership to Advance Throughput Computing (PATh),
Advanced Cyberinfrastructure Coordination Ecosystem: Services {\&} Support (ACCESS),
Frontera and Ranch computing project at the Texas Advanced Computing Center,
U.S. Department of Energy-National Energy Research Scientific Computing Center,
Particle astrophysics research computing center at the University of Maryland,
Institute for Cyber-Enabled Research at Michigan State University,
Astroparticle physics computational facility at Marquette University,
NVIDIA Corporation,
and Google Cloud Platform;
Belgium {\textendash} Funds for Scientific Research (FRS-FNRS and FWO),
FWO Odysseus and Big Science programmes,
and Belgian Federal Science Policy Office (Belspo);
Germany {\textendash} Bundesministerium f{\"u}r Forschung, Technologie und Raumfahrt (BMFTR),
Deutsche Forschungsgemeinschaft (DFG),
Helmholtz Alliance for Astroparticle Physics (HAP),
Initiative and Networking Fund of the Helmholtz Association,
Deutsches Elektronen Synchrotron (DESY),
and High Performance Computing cluster of the RWTH Aachen;
Sweden {\textendash} Swedish Research Council,
Swedish Polar Research Secretariat,
Swedish National Infrastructure for Computing (SNIC),
and Knut and Alice Wallenberg Foundation;
European Union {\textendash} EGI Advanced Computing for research;
Australia {\textendash} Australian Research Council;
Canada {\textendash} Natural Sciences and Engineering Research Council of Canada,
Calcul Qu{\'e}bec, Compute Ontario, Canada Foundation for Innovation, WestGrid, and Digital Research Alliance of Canada;
Denmark {\textendash} Villum Fonden, Carlsberg Foundation, and European Commission;
New Zealand {\textendash} Marsden Fund;
Japan {\textendash} Japan Society for Promotion of Science (JSPS)
and Institute for Global Prominent Research (IGPR) of Chiba University;
Korea {\textendash} National Research Foundation of Korea (NRF);
Switzerland {\textendash} Swiss National Science Foundation (SNSF).

\end{document}